\documentclass[aps,showpacs,twocolumn]{revtex4}
\usepackage{graphicx}

\begin{document}

\title{Cosmology and two-body problem of D-branes}

\author{Keitaro Takahashi$^{(1)}$ and Kazuhide Ichikawa$^{(2)}$}


\affiliation{$^{(1)}$Department of Physics, University of Tokyo, 
7-3-1 Hongo, Bunkyo, Tokyo 113-0033, Japan}

\affiliation{$^{(2)}$Research Center for the Earth Universe, 
University of Tokyo, 7-3-1 Hongo, Bunkyo, Tokyo 113-0033, Japan}

\date{\today}

\begin{abstract}
In this paper, we investigate the dynamics and the evolution of 
the scale factor of a probe Dp-brane which move in the background of 
source Dp-branes. Action of the probe brane is described by the Born-Infeld 
action and the interaction with the background R-R field.
When the probe brane moves away from the source branes, it expands
by power law, whose index depends on the dimension of the brane. 
If the energy density of the gauge field on the brane 
is subdominant, the expansion is decelerating irrespective of 
the dimension of the brane. On the other hand, when the probe brane
is a Nambu-Goto brane, the energy density of the gauge field
can be dominant, in which case accelerating expansion occurs for $p \leq 4$. 
The accelerating expansion stops when the brane has expanded sufficiently 
so that the energy density of the gauge field become subdominant.

\end{abstract}

\pacs{98.80.Cq  04.50.+h  11.25.Wx}

\maketitle


\section{Introduction}

By the discovery of D-brane, not only string theory but also
cosmology have been activated significantly. The Randall-Sundrum braneworld 
model \cite{RS1,RS2,RS3} is the simplest cosmological model which was
induced by the idea of D-brane. In this model, the action of the brane is 
assumed to be the Nambu-Goto action. Cosmology with Born-Infeld action has
also been investigated in \cite{DBC1,DBC2,DBC3} and it is found that 
behavior of the gauge field confined to the brane is significantly different 
from that of a gauge field added to the Nambu-Goto brane. Interaction 
between D-branes by R-R charge, which is absence in the Randall-Sundrum model,
has been studied as a potential energy source that inflate the brane 
by many authors \cite{DBI1,DBI2,DBI3,DBI4,DBI5,DBI6,DBI7,DBI8,DBI9,DBI10}. 
For a review of cosmology in the context of string theory, see, for example, 
\cite{DBI_review}.

Since D-brane is a fundamental object in superstring theory,
their two-body problem is also fundamental.
Burgess et. al. \cite{Branonium} studied motion of a probe brane 
in the background spacetime of source branes and found that there exist 
bound states of D6-brane and anti-D6-brane, which they called 'branonium'.
Probe-brane dynamics was also discussed in \cite{PBD,Branonium2}.
Recently cosmology on the probe brane was studied in the context of
bouncing universe \cite{Branonium3}.

In this paper, we investigate two-body problem and cosmology of D-branes. 
Basic approach is the same as \cite{Branonium,Branonium3} but we take 
into account a gauge field confined to the probe brane, which was neglected 
in \cite{Branonium,Branonium3}. The motion of the brane causes the time
evolution of the induced metric on it, which is seen as cosmological
expansion or contraction by an observer living on the brane. In this sense, 
our picture is similar to that of 'mirage cosmology' 
\cite{mirage1,mirage2,mirage3,mirage4}. 
Thus, by following the motion of the brane, 
we can also follow the evolution of the scale factor. 
We show that the gauge field on the probe brane, 
which has not been studied rigorously, can affect the behavior of 
the scale factor.

The rest of this paper is organized as follows. 
In section \ref{section:background}, we review the $p$-brane solutions
in supergravity as the background spacetime of the source D-branes.
We consider the motion of a probe brane in this background spacetime
in section \ref{section:dynamics} and follow the evolution of the scale
factor on the probe brane in section \ref{section:cosmology}.
In section \ref{section:discussion} and \ref{section:summary},
we give discussion and summary, respectively.

\section{Background Spacetime \label{section:background}}

We consider a system in which a probe D-brane (or anti-D-brane) moves
within the background of $N$ parallel source D-branes.
In this section, we review the p-brane solutions in supergravity
as the background spacetime of the source D-branes.
Low-energy effective theories for superstring theories are given by
supergravities, among which we consider only Type IIA and IIB
here for simplicity. The effective actions include the metric,
the 2-form potential and the scalar dilaton in the NS-NS sector,
$(n-1)$-form gauge potentials in the R-R sector and Chern-Simons terms. 
Here $n$ is even for IIA and odd for IIB.

To obtain a tractable system to study, we shall make 
a {\it consistent truncation} (see \cite{Stelle98} and references therein) 
of the action down to a simple system comprising only the metric $G_{MN}$,
the scalar dilaton $\phi$ and a single $(n-1)$-form gauge potential 
$A_{[n-1]}$ with corresponding field strength $F_{[n]}$.
Then the background spacetime of source D$p$-brane are determined 
by the following action in the Einstein Frame,
\begin{equation}
S = \int D^{D} x \sqrt{-G} 
         \left[ R - \frac{1}{2} \partial^{M} \phi \partial_{M} \phi
                  - \frac{1}{2 n!} e^{a \phi} F^{2}_{[n]}
         \right],
\end{equation}
where $D = 10$ and $a = (5-n)/2$ is the dilaton coupling of the R-R field.
Assuming asymptotic flatness and spherical symmetry in the transverse 
directions, flatness of the branes and an ``electric'' gauge field, 
the background spacetime and gauge field for $p \leq 6$ are given by,
\begin{equation}
ds^{2} = h^{-(7-p)/8} \eta_{\mu\nu} dx^{\mu} dx^{\nu}
       + h^{(p+1)/8} \delta_{mn} dy^{m} dy^{n},
\end{equation}
\begin{equation}
e^{\phi} = h^{(3-p)/4},
\label{eq:dilaton}
\end{equation}
\begin{equation}
A_{M_{1} M_{2} \cdots M_{p+1}} =
               \epsilon_{M_{1} M_{2} \cdots M_{p+1}} (1 - h^{-1}),
\end{equation}
where $x^{\mu} (\mu=0,1,\cdots,p)$ and $y^{m} (m=1,2,\cdots,D-p-1)$ are 
the coordinates parallel and transverse to the branes, respectively. 
We define the radial coordinate transverse to the brane as 
$r^{2} \equiv \delta_{mn} y^{m} y^{n}$ and then,
\begin{equation}
h(r) = 1 + \frac{k}{r^{7-p}}.
\end{equation}
Here $k$ is an integration constant which represent the energy scale of
the source branes:
\begin{equation}
k = (2 \sqrt{\pi})^{5-p} \Gamma(\frac{7-p}{2}) g_{\rm s} l_{\rm s}^{7-p} N,
\end{equation}
where $g_{\rm s}$ is the string coupling constant at infinity, 
$l_{\rm s}$ is the string length scale and $N$ is the number of the
source branes. It should be noted that these solutions are reliable
only for $r \gg l_{\rm s}$. This is because supergravity is a good 
approximation of superstring theory only within this region, where
the brane interactions are dominated by massless string states.

The asymptotic behaviors of the gravitational field and the gauge field
potential can be understood in terms of Gauss' law. Both behave asymptotically
like $\sim r^{-(7-p)}$, as expected from the Laplace equation,
\begin{equation}
\nabla^{2} f(r) = 
\left[ \frac{d^{2}}{dr^{2}} + \frac{8-p}{r} \frac{d}{dr} \right] f(r)
= 0.
\end{equation}
Thus, the potential produced by a D6-brane is, like that of a point 
particle in ordinary 4-dimensional spacetime, $\sim r^{-1}$.
For global structures of these solutions, see, for example, \cite{Peet00}.

On the other hand, there is no asymptotically-flat solution for $p \geq 7$.
Hereafter we concentrate on $p \leq 6$ cases.

\section{Dynamics of Probe Brane \label{section:dynamics}}

In this section we consider the motion of a probe brane, which is assumed
to be parallel to the source branes, in the background spacetime
discussed in the previous section. The dynamics of the probe brane
which has ``electric'' charge is determined by the Born-Infeld action 
(in the String Frame),
\begin{equation}
S_{\rm BI} = - T_{p} \int d^{p+1} x e^{- \phi}
             \sqrt{- \det{(g_{\mu\nu} + 2 l_{\rm s}^{2} F_{\mu\nu})}}
\end{equation}
and the interaction with the background gauge field $A_{[p+1]}$,
\begin{equation}
S_{\rm WZ} = - q T_{p} \int A_{[p+1]}.
\end{equation}
Here $g_{\mu\nu}$ is the induced metric on the probe brane,
$F_{\mu\nu}$ is the $U(1)$ gauge field strength confined to the brane
and $q$ is the R-R charge of the brane, which equals to $\pm 1$ for
D-brane and anti-D-brane, respectively. Note that the field strength 
$F_{\mu\nu}$ should be thermal in nature in order not to break the isotropy 
of the brane. Therefore, we interpret that
$F_{\mu\nu}F^{\mu\nu} \rightarrow \langle F_{\mu\nu}F^{\mu\nu} \rangle$ etc
\cite{D-brane_review}. The induced metric on the brane
in the String Frame is written as,
\begin{eqnarray}
d\tilde{s}^{2} 
& = & e^{4 \phi/(D-2)} ds^{2} \nonumber \\
& = & - h^{-1/2} (1 - h v^{2}) dt^{2} 
  + h^{-1/2} \delta_{ij} dx^{i} dx^{j}
\label{eq:induced-metric}
\end{eqnarray}
where we took the static gauge, $t = x^{0}$ and $i,j = 1,2,\cdots,p$. 
Here we defined the velocity $v$ of the brane as,
\begin{equation}
v^{2} \equiv \delta_{mn} \frac{dy^{m}}{dt} \frac{dy^{n}}{dt}.
\end{equation}
Thus the motion of the brane in the dimensions transverse to the brane 
is described in terms of the radial coordinate $r$ and the velocity $v$.

Due to the spherical symmetry in the transverse direction,
the angular momenta of the brane are conserved. This shows that the motion
is confined to the plane which is spanned by the initial position and
momentum vectors. We will denote the polar coordinate in this plane by
$r$ and $\theta$. Further, due to technical difficuluty, we treat the gauge
field as a perturbation and consider the leading term. 
Then the total Lagrangian of the probe brane is,
\begin{equation}
L = - m h^{-1}
\left[ \sqrt{1 - h (\dot{r}^{2} + r^{2} \dot{\theta}^{2})} 
       (1 + l_{\rm s}^{4} F_{\mu\nu}F^{\mu\nu}) - q 
\right],
\label{eq:Lagrangian}
\end{equation}
where we have neglected an additive constant, $m = T_{p} \int d^{p} x$ is 
the ``mass'' of the brane, and dot denotes a derivative with respect to $t$. 
The independent variables are $r, \theta$ and the gauge potential $A_{\mu}$ 
on the brane. The canonical momenta associated with these variables are,
\begin{eqnarray}
p_{r} & \equiv & m^{-1} \frac{\partial L}{\partial \dot{r}} \nonumber \\
& = & \frac{\dot{r}}{\sqrt{1 - h (\dot{r}^{2} + r^{2} \dot{\theta}^{2})}}
      (1 + l_{\rm s}^{4} F_{\mu\nu}F^{\mu\nu}), 
\label{eq:momentum} \\
l & \equiv & m^{-1} \frac{\partial L}{\partial \dot{\theta}} \nonumber \\
& = & \frac{r^{2} \dot{\theta}}{\sqrt{1 - h (\dot{r}^{2} + r^{2} \dot{\theta}^{2})}}
      (1 + l_{\rm s}^{4} F_{\mu\nu}F^{\mu\nu}), 
\label{eq:a-momentum} \\
p_{A}^{i} & \equiv & m^{-1} \frac{\partial L}{\partial \dot{A}_{i}} 
\nonumber \\
& = & - \frac{4 \sqrt{1 - h (\dot{r}^{2} + r^{2} \dot{\theta}^{2})}}{h} F^{i0},
\end{eqnarray}
where $p_{A}^{i}$ is also conserved as we can see from the Euler-Lagrange
equation. Thus the ``electric field'' $F^{i0}$
can be written in terms of the other variables. On the other hand,
the ``magnetic field'' $F^{ij}$ are obtained from Bianchi identity,
\begin{equation}
\partial_{\mu} F_{\nu\lambda} + \partial_{\nu} F_{\lambda\mu} 
+ \partial_{\lambda} F_{\mu\nu} = 0,
\end{equation}
as,
\begin{equation}
F_{ij} = C_{ij} = {\rm const}.
\end{equation}
Combining the above results, it follows that,
\begin{eqnarray}
F_{\mu\nu} F^{\mu\nu} & = &
\left( \delta^{ik} \delta^{jl} C_{ij} C_{kl} 
       - \frac{\delta_{ij} p_{A}^{i} p_{A}^{j}}{8} \right) h \nonumber \\
& \equiv & C' h,
\end{eqnarray}
where $C'$ is a constant which represents the energy scale of
the gauge field.

From Eq. (\ref{eq:momentum}) and (\ref{eq:a-momentum}), 
we obtain the following useful relation:
\begin{equation}
\dot{r}^{2} + r^{2} \dot{\theta}^{2}
= \frac{p_{r}^{2} + l^{2}/r^{2}}
       {(1 + C h)^{2} + h (p_{r}^{2} + l^{2}/r^{2})},
\label{eq:useful}
\end{equation}
where $C \equiv C' l_{\rm s}^{4}$ is a dimensionless constant which represents
the energy scale of the gauge field in units of $l_{\rm s}^{-1}$.
Then the Hamiltonian can be written as,
\begin{eqnarray}
E & \equiv & p_{r} \dot{r} + l \dot{\theta} + p_{A}^{i} \dot{A}_{i} - m^{-1} L 
\nonumber \\
& = & \frac{ \left\{1 + (4 D + C)h \right\}(1 + C h) 
             + h (p_{r}^{2} + l^{2}/r^{2})}
           {h \sqrt{(1 + C h)^{2} + h (p_{r}^{2} + l^{2}/r^{2})}} \nonumber \\
&   & - \frac{q}{h},
\label{eq:energy}
\end{eqnarray}
which gives the conserved energy. Here we took a gauge $A_{0} = 0$
and $D \equiv \delta_{ij} p_{A}^{i} p_{A}^{j} /16$.
Note that this agree with (2.22) of \cite{Branonium} in the limit of 
no gauge field, $C,D \rightarrow 0$. Hereafter, we set $D = 0$ for simplicity,
which means that there is only magnetic field. From Eq. (\ref{eq:energy}),
we expect that the dynamics does not change so much even if there
are both electric and magnetic fields.

Following \cite{Branonium}, we define effective potential $V_{\rm eff}$ 
for the radial motion as,
\begin{eqnarray}
V_{\rm eff}(r) & \equiv & E(p_{r} = 0) \nonumber \\
& = & h^{-1} \left[ \sqrt{(1 + C h)^{2} + h l^{2}/r^{2}} - q \right].
\end{eqnarray}
The asymptotic behavior depends on the charge and the dimension of the brane.
For $p=6$,
\begin{equation}
V_{\rm eff}(r) \rightarrow 
\left\{ \begin{array}{ll}
\frac{l}{\sqrt{k}} r^{-1/2} & \mbox{for}\ r \rightarrow 0 \\
1+C-q + k(q-1)r^{-1} & \\
\;\;\; + \left[ \frac{l^{2}}{2(1+C)} - k(q-1) \right] r^{-2} & 
\mbox{for}\ r \rightarrow \infty
\end{array} \right.
\end{equation}
For $p=5$,
\begin{equation}
V_{\rm eff}(r) \rightarrow 
\left\{ \begin{array}{l}
\frac{l}{\sqrt{k}} 
- \left( \frac{l^{2}}{2 k \sqrt{k}} + \frac{q}{k} \right) r^{2}
\hspace{1cm} \mbox{for}\ r \rightarrow 0 \\ 
1 + C - q + \left[ \frac{l^{2}}{2(1+C)} + k(q-1) \right] r^{-2} \\ 
\hspace{4cm} \mbox{for}\ r \rightarrow \infty
\end{array} \right.
\end{equation}
For $p\leq4$,
\begin{equation}
V_{\rm eff}(r) \rightarrow 
\left\{ \begin{array}{ll}
\frac{l}{\sqrt{k}} r^{(5-p)/2} & 
\mbox{for}\ r \rightarrow 0 \\ 
1 + C - q + \frac{l^{2}}{2(1+C)} r^{-2} & 
\mbox{for}\ r \rightarrow \infty
\end{array} \right.
\end{equation}
As is pointed out in \cite{Branonium}, there exist stable bound orbits
in the case of anti-$6$-brane.

\begin{figure*}
\includegraphics{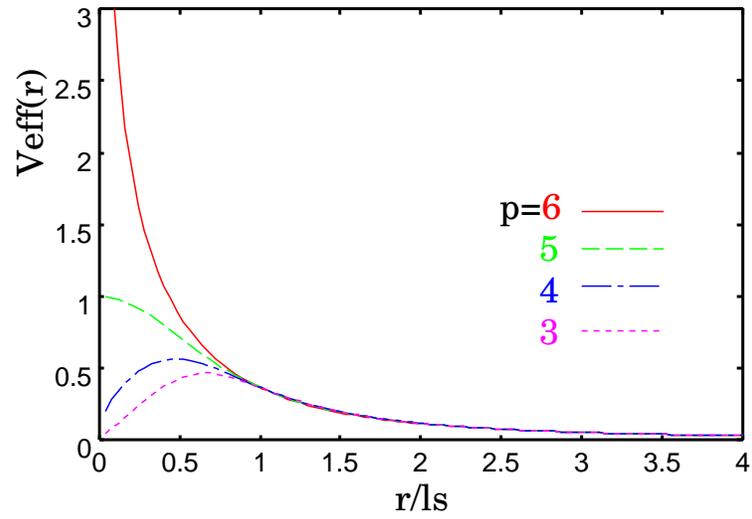}
\caption{Effective potential $V_{\rm eff}$ for the radial motion of
the probe brane, varying its spatial dimension $p$.
Other parameters are set as $k=l=1$ and $C=0$.
\label{fig:Veff-p}}
\end{figure*}

\begin{figure*}
\includegraphics{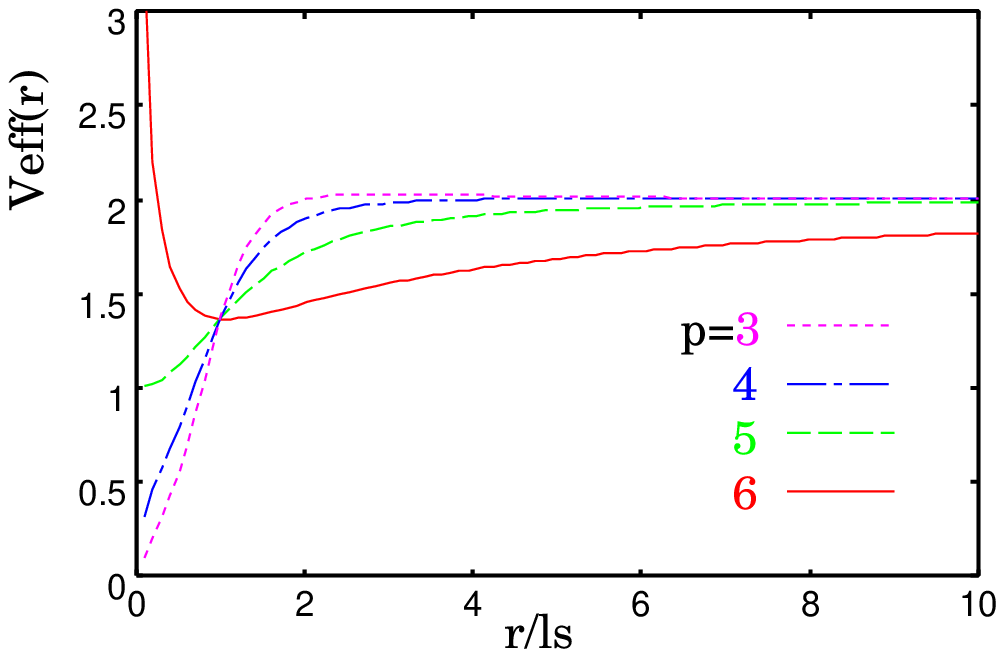}
\caption{Effective potential $V_{\rm eff}$ for the radial motion of
the probe anti-brane, varying its spatial dimension $p$.
Other parameters are set as $k=l=1$ and $C=0$.
\label{fig:Veff-p-anti}}
\end{figure*}

\begin{figure*}
\includegraphics{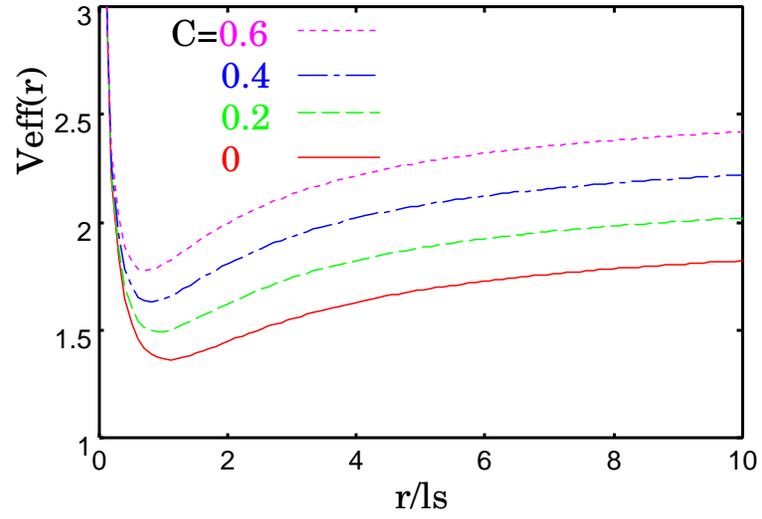}
\caption{Effective potential $V_{\rm eff}$ for the radial motion of
the probe anti-6-brane, varying the energy scale $C$ of the gauge field
on it. Other parameters are set as $k=l=1$.
\label{fig:Veff-C-anti}}
\end{figure*}

Behavior of the effective potential is shown in Fig. \ref{fig:Veff-p},
\ref{fig:Veff-p-anti} and \ref{fig:Veff-C-anti}. Fig. \ref{fig:Veff-p}
and \ref{fig:Veff-p-anti} shows the effective potential of $p$-brane 
and anti-$p$-brane for various $p$, respectively. From this, we can see
that there can be stable bound state in the case of anti-$6$-brane,
as is expected. Note that the position of the potential minimum,
$r_{\rm min}$, depends on the angular momentum $l$ and $r_{\rm min}$
can be much larger than $l_{\rm s}$ if $l$ is sufficiently large.
Fig. \ref{fig:Veff-C-anti} shows the effective potential of $6$-brane 
for various $C$. As can be seen, qualitative feature does not depend on $C$.

Using Eq. (\ref{eq:momentum}), (\ref{eq:a-momentum}) and (\ref{eq:energy}),
$\dot{r}$ can be expressed in terms of $r, E, l$:
\begin{equation}
\dot{r}^{2} = h^{-1}
\left[
1 - \frac{r^{2} (1 + C h)^{2} + h l^{2}}{r^{2} (E h + q)^{2}}
\right].
\label{eq:r-dot}
\end{equation}
We can follow the motion of the brane by integrating this equation.
Since, as can be seen from Eq. \ref{eq:induced-metric}, the scale factor
on the brane is a function of $r$, its evolution can also be calculated
from this equation as we discuss in the next subsection.
Note that this equation corresponds to Friedmann equation and that
this reduces to the Friedmann equation in \cite{Branonium3}
in the limit of $C \rightarrow 0$.

The brane trajectory can be calculated as follows: 
define $u \equiv 1/r$ and then,
\begin{equation}
u' \equiv \frac{du}{d\theta} = - r^{-2} \frac{dr}{d\theta} 
   = - r^{-2} \frac{\dot{r}}{\dot{\theta}} = - \frac{p_{r}}{l}.
\end{equation}
Eliminating $p_{r}$ from Eq. (\ref{eq:energy}) using this equation,
we obtain,
\begin{equation}
E = h^{-1} \left[ \sqrt{(1+Ch)^{2} + h l^{2} (u^{2} + u'^{2})} - q \right],
\end{equation}
from which the orbit is obtained as,
\begin{equation}
\theta - \theta_{0} =
\int^{1/r}_{1/r_{0}} \frac{du}{\sqrt{A + B u^{7-p} - u^{2}}},
\end{equation}
where,
\begin{eqnarray}
A & = & l^{-2} (E^{2} + 2 E q - C^{2} - 2 C) \\
B & = & E^{2} - C^{2}.
\end{eqnarray}
Thus the orbit of the probe brane is equivalent to that of a classical
nonrelativistic particle in the central potential proportional to $r^{p-7}$,
even when there exists a gauge field on the brane. Especially for $p = 6$,
the bound orbit is closed.

\section{Cosmology on Probe Brane \label{section:cosmology}}

\subsection{Evolution of Scale Factor}

From the induced metric on the brane Eq. (\ref{eq:induced-metric}), 
the scale factor $a$ is given by,
\begin{equation}
a = h^{-1/4}.
\label{eq:scale-factor}
\end{equation}
On the other hand, the cosmological time $\tau$ on the brane is expressed as,
\begin{eqnarray}
\tau & \equiv & \int h^{-1/4} 
                     \sqrt{1 - h (\dot{r}^{2} + r^{2} \dot{\theta}^{2})} dt 
                \nonumber \\
& = & \int h^{-1/4} \frac{1 + C h}{E h + q} \dot{r}^{-1} dr \nonumber \\
& = & \int h^{1/4} 
           \frac{1 + C h}{\sqrt{(E h + q)^{2} - (1 + C h)^{2} - h l^{2}/r^{2}}}
           dr.
\label{eq:tau}
\end{eqnarray}
Here we used Eq. (\ref{eq:useful}) and (\ref{eq:energy}) in the
second equation and Eq. (\ref{eq:r-dot}) in the last equation. 
Thus, from Eq. (\ref{eq:scale-factor}) and (\ref{eq:tau}),
the scale factor $a$ can be obtained as a function of $\tau$.

Here we define two characteristic radii: gravitational radius $r_{\rm g}$
and gauge-field radius $r_{c}$. The former corresponds to the
Schwarzshild radius,
\begin{equation}
r_{\rm g} \equiv k^{1/(7-p)}.
\end{equation}
It should be noted that,
\begin{equation}
h(r) \approx \left\{ \begin{array}{ll}
             k/r^{7-p} & \mbox{for}\ r \ll r_{\rm g} \\
             1         & \mbox{for}\ r \gg r_{\rm g}.
              \end{array} \right.
\end{equation}
The latter represents the radius, below which the approximation of 
the Lagrangian (\ref{eq:Lagrangian}) breaks down 
($l_{\rm s}^{4} F_{\mu\nu}F^{\mu\nu} = C h(r_{\rm c}) = 1$):
\begin{eqnarray}
r_{\rm c} & \equiv & \left( \frac{C k}{1 - C} \right)^{1/(7-p)} \nonumber \\
      & \approx & (C k)^{1/(7-p)}.
\end{eqnarray}
Hereafter we consider the case $r_{\rm g} \gg l_{\rm s}$ because otherwise
the scale factor does not change so much in the region where the background
solution is reliable ($r \gg l_{\rm s}$).

Then let us consider a situation that the probe brane goes away from
the neighborhood of the source branes 
(but $r = r_{0} \gg l_{\rm s}, r_{\rm c}$, of cource) to infinity.
When $r \ll r_{\rm g}$, the relation between the scale factor and
the cosmological time is simple. In this case Eq. (\ref{eq:tau}) becomes,
noting that $h \gg 1$ and $C h < 1$,
\begin{eqnarray}
\tau & \approx & \int_{r_{0}}^{r} h^{1/4} \frac{1 + C h}{E h} dr \nonumber \\
     & \approx & E^{-1} k^{-3/4} 
                 \int_{r_{0}}^{r} r^{3(7-p)/4} (1 + C k r^{p-7}) dr.
\end{eqnarray}
Note that this is independent on $q$ in this limit.
First we consider the case $C = 0$. Then,
\begin{equation}
\tau = \frac{4}{25 - 3 p} E^{-1} k^{-3/4} 
           (r^{(25-3p)/4} - r_{0}^{(25-3p)/4}). 
\end{equation}
At late time ($r \gg r_{0}$), we obtain,
\begin{equation}
\tau \propto r^{(25-3p)/4},
\end{equation}
from which the evolution of the scale factor is obtained as,
\begin{eqnarray}
a(\tau) & = & h^{-1/4} \approx (k r^{p-7})^{-1/4} \nonumber \\
        & \propto & \tau^{(7-p)/(25-3p)}.
\end{eqnarray}
Here $(7-p)/(25-3p) = 1/7, 1/5, 3/13, 1/4, 5/19, 3/11$ for 
$p = 6, 5, \cdots, 1$. Although the expansion becomes faster
with smaller $p$, acceleration phase cannot be realized.

If $C \not= 0$, a correction term is added,
\begin{equation}
a(\tau) \propto (\tau^{(7-p)/(25-3p)} - A C a_{1}(\tau)),
\end{equation}
where $A$ is a constant which depends on $E, k, p$ and $a_{1}$ is,
to leading term,
\begin{equation}
a_{1}(\tau) = 
\left\{ \begin{array}{ll}
\tau^{-3(7-p)/(25-3p)}
& \mbox{for}\ p \geq 4 \\
\tau^{-3/4} \log \tau
& \mbox{for}\ p = 3 \\
\tau^{-1} (r_{0}^{-(3-p)/4} - B \tau^{-16/(3-p)(25-3p)})
& \mbox{for}\ p \leq 2,
\end{array} \right.
\end{equation}
where $B$ is also a constant which depends on $E, k, p$.
It should be noted that the effect of the gauge field decrease as
the brane expands since its energy density decreases as $h \propto a^{-4}$.

\begin{figure*}
\includegraphics{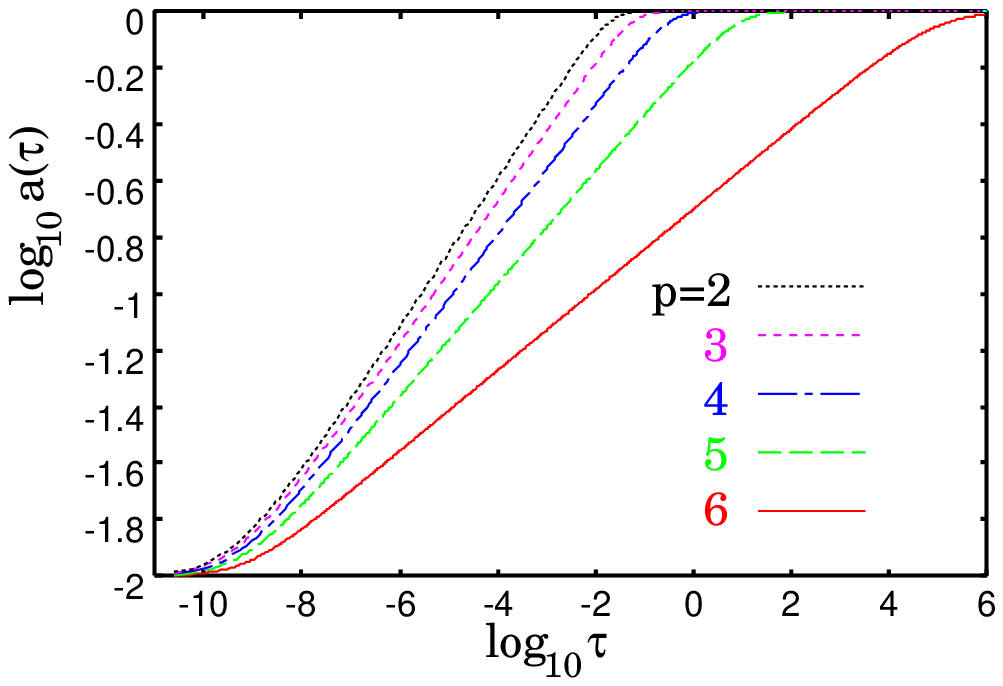}
\caption{Evolution of the scale factor $a(\tau)$ without the gauge field
on the brane for various $p$. Other parameters are set as, 
$k=10^{8}, E=10^{3}, l=10, q=-1, r_{0}=1$.
\label{fig:scale_no-gauge}}
\end{figure*}

\begin{figure*}
\includegraphics{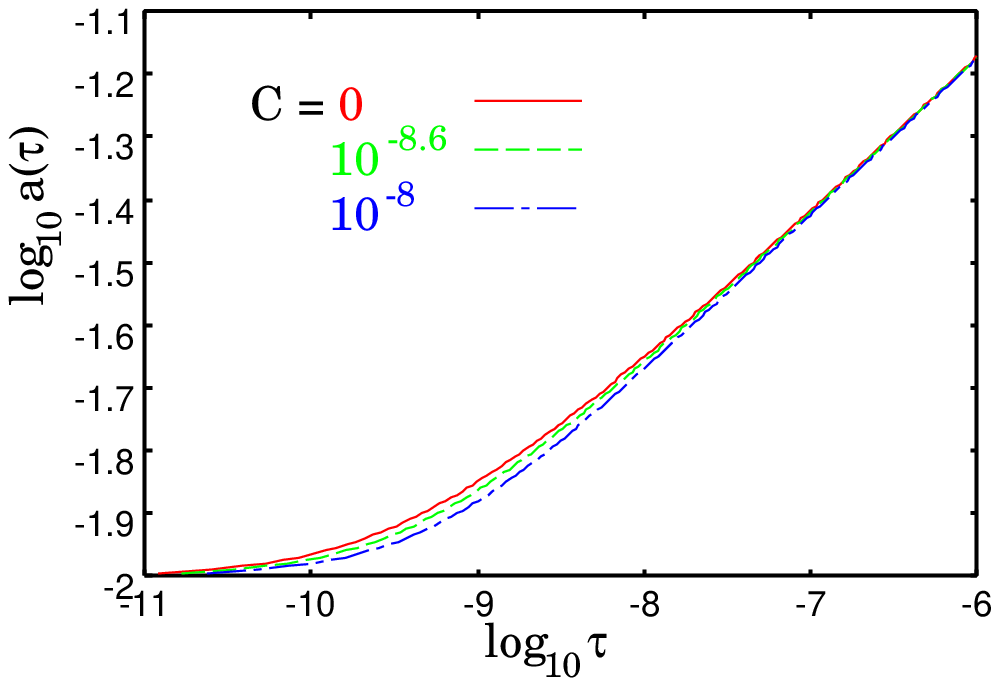}
\caption{Evolution of the scale factor $a(\tau)$ with the gauge field
on the brane for various $C$. Other parameters are set as, 
$p=4, k=10^{8}, E=10^{3}, l=10, q=-1, r_{0}=1$.
\label{fig:scale_gauge}}
\end{figure*}

When $r$ becomes much larger than $r_{\rm g}$, the scale factor
stops to evolve and becomes almost unity. The behavior of the scale factor
in the case of no gauge field is shown in Fig. \ref{fig:scale_no-gauge}.
As is expected, the scale factor evolves by power-law and then
decelerate quickly to become unity. Fig. \ref{fig:scale_gauge} shows
the effect of the gauge field on the brane. As can be seen, the effect
is very small even if $C$ is as large as possible and the late-time
behavior is independent on the existence of the gauge field.

We can also see the evolution of the scale factor by the effective
Friedmann equation which can be derived from Eq. (\ref{eq:r-dot}):
\begin{eqnarray}
a^{-2} \left( \frac{da}{d \tau} \right)^{2} & = &
\frac{(7-p)^{2}}{16} k^{-\frac{2}{7-p}} a^{-\frac{2(11-p)}{7-p}}
(1 - a^{4})^{\frac{2(8-p)}{7-p}} \nonumber \\
& & \times (1 + C a^{-4})^{-2}
\nonumber \\
& & \times 
\left[
(E^{2} - C^{2}) a^{-4} + 2(Eq - C) 
\right.\nonumber \\
& & 
\left. \;\;\;\;\;\;
- l^{2} k^{-\frac{2}{7-p}} a^{-\frac{8}{7-p}} (1 - a^{4})^{\frac{2}{7-p}}
\right],
\end{eqnarray}
which agrees with \cite{mirage1} in the limit of $C \rightarrow 0$.

\subsection{High Energy Limit}

\begin{figure*}
\includegraphics{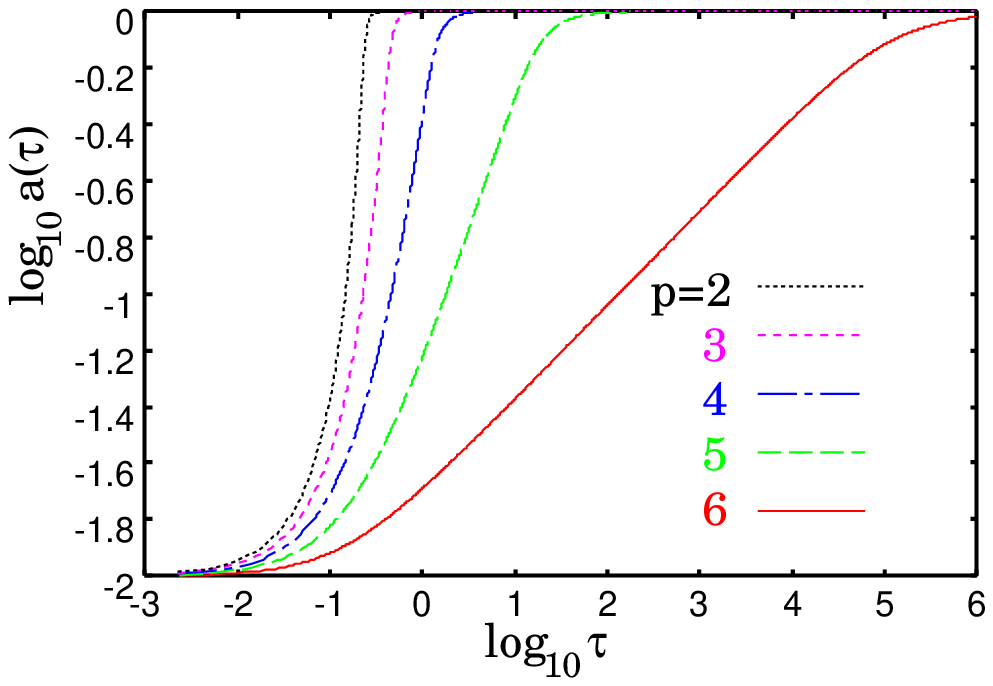}
\caption{Evolution of the scale factor $a(\tau)$ of the brane dominated by
the gauge field for various $p$. Other parameters are set as, 
$k=10^{8}, E=10^{3}, l=10, q=-1, r_{0}=1, C=1$.
\label{fig:scale_high-gauge}}
\end{figure*}

Here we consider the probe brane to be a Nambu-Goto brane with gauge field
and the same R-R charge as a D-brane, 
for which the Lagrangian (\ref{eq:Lagrangian}) is exact. In this case, we can
take a high-energy limit ($C h \gg 1$). Although, for a D-brane, this limit 
is in contradiction to the approximation which we used to derive
(\ref{eq:Lagrangian}), we could still obtain tendency of high-energy effect,
as is often done in higher derivative theory.
In this regime, Eq. (\ref{eq:tau}) for $r \ll r_{\rm g}$ becomes,
\begin{eqnarray}
\tau & \approx & \int_{r_{0}}^{r} h^{1/4} \frac{C}{\sqrt{E^{2} - C^{2}}} dr 
                 \nonumber \\
     & \approx & \frac{C k^{1/4}}{\sqrt{E^{2} - C^{2}}} 
                 \int_{r_{0}}^{r} r^{(p-7)/4} dr.
\end{eqnarray}
For $p \geq 4$,
\begin{eqnarray}
\tau & = & \frac{4}{p-3} \frac{C k^{1/4}}{\sqrt{E^{2} - C^{2}}}
           (r^{(p-3)/4} - r_{0}^{(p-3)/4}) \nonumber \\
& \stackrel{r \gg r_{0}}{\longrightarrow} & \propto r^{(p-3)/4},
\end{eqnarray}
then,
\begin{equation}
a(\tau) \propto \tau^{(7-p)/(p-3)},
\end{equation}
where $(7-p)/(p-3) = 1/3, 1, 3$ for $p = 6, 5, 4$.
Thus accelerating expansion is realized for $p = 4$. Next, for $p = 3$,
\begin{equation}
\tau = \frac{C k^{1/4}}{\sqrt{E^{2} - C^{2}}} \log{\frac{r}{r_{0}}},
\end{equation}
then,
\begin{equation}
a(\tau) = k^{-1/4} r_{0} 
          \exp{\left( 
               \frac{\sqrt{E^{2} - C^{2}}}{C k^{1/4}} \tau
               \right)}.
\end{equation}
Thus the scale factor increase exponentially. Finally for $p \leq 2$,
\begin{equation}
\tau = \frac{4}{3-p} \frac{C k^{1/4}}{\sqrt{E^{2} - C^{2}}}
       (r_{0}^{-(3-p)/4} - r^{-(3-p)/4}),
\end{equation}
then,
\begin{equation}
a(\tau) = k^{-1/4}
          \left( r_{0}^{-(3-p)/4} - 
                 \frac{3-p}{4} \frac{\sqrt{E^{2} - C^{2}}}{C k^{1/4}} \tau
          \right)^{-\frac{7-p}{3-p}}.
\end{equation}
It can be easily shown that the expansion is accelerating in this case.
These analyses are confirmed in Fig. \ref{fig:scale_high-gauge}.

\begin{figure*}
\includegraphics{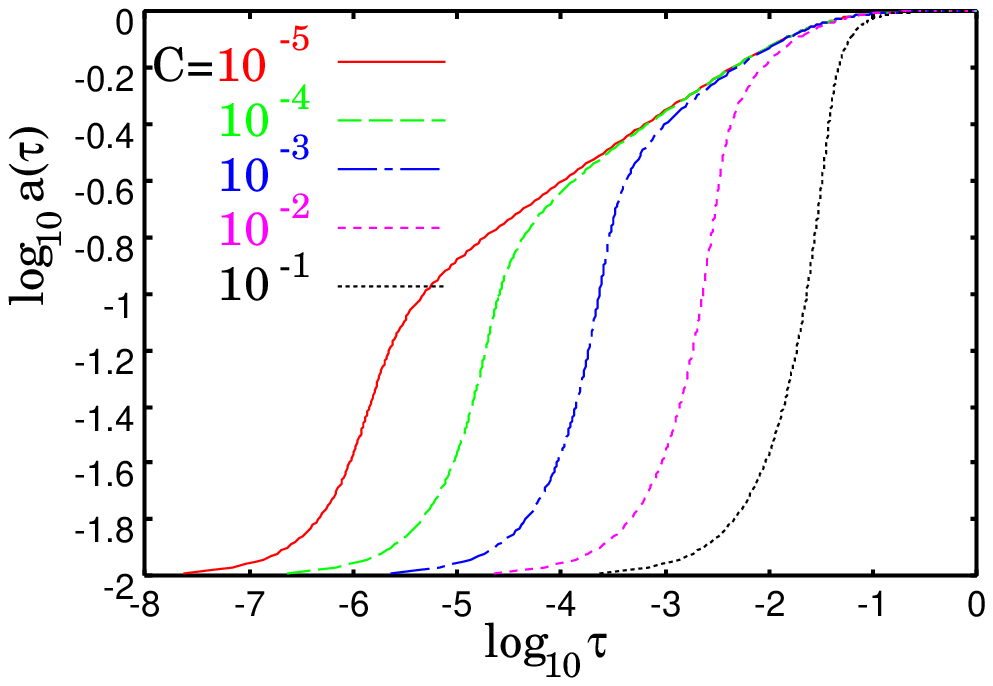}
\caption{Evolution of the scale factor $a(\tau)$ of the brane dominated by
the gauge field for various $C$. Other parameters are set as, 
$p=3, k=10^{8}, E=10^{3}, l=10, q=-1, r_{0}=1$.
\label{fig:scale_high-gauge_C}}
\end{figure*}

As stated in the previous subsection, the energy density of the gauge
field decrease as the brane expands. With the parametrization in 
Fig. \ref{fig:scale_high-gauge}, the gauge field is dominant for
all over the evolution since $C$ is sufficiently large so that $C h$ 
at infinity is still large ($C h(r = \infty) = C = 1$). 
If $C$ is smaller than unity, late phase will behave like that of 
the case discussed in the previous subsection, even if accelerating 
expansion occurs in early phase. In Fig. \ref{fig:scale_high-gauge_C}, 
we show the cases with intermediate $C$. We can see the transition
from accelerating phase to decelerating phase. Of course, the transition
occurs earlier with smaller $C$.

\subsection{Einstein Frame}

Finally, we give the evolution of the scale factor in the Einstein Frame.
The procedure is almost the same as in the String Frame. The induced metric
in the Einstein Frame is,
\begin{equation}
ds^{2} = - h^{-(7-p)/8} (1 - h v^{2}) dt^{2}
         + h^{-(7-p)/8} \delta_{ij} dx^{i} dx^{j}.
\end{equation}
Then the cosmological time is,
\begin{equation}
\tau = \int_{r_{0}}^{r} h^{-(7-p)/16} \sqrt{1 - h v^{2}} dr.
\end{equation}
With $C = 0$, the scale factor evolves as, for $r \gg r_{\rm g}$,
\begin{equation}
a(\tau) = h^{-(7-p)/16} \propto \tau^{(7-p)^{2}/(11-p)^{2}},
\end{equation}
where the index is $(7-p)^{2}/(11-p)^{2} = 1/25, 1/9, 9/49, 1/4, 25/81, 9/25$ 
for $p = 6, 5, \cdots, 1$. In high-energy limit ($C h \gg 1$),
for $p \not= 3$,
\begin{equation}
a(\tau) \propto t^{(7-p)^{2}/(3-p)^{2}},
\end{equation}
where $(7-p)^{2}/(3-p)^{2} = 1/9, 1, 9, 25, 9$ for $p = 6, 5, 4, 2, 1$.
For $p = 3$,
\begin{equation}
a(\tau) \propto \exp{\left(
                     \frac{\sqrt{E^{2} - C^{2}}}{C k^{1/4}} \tau
                     \right)}.
\end{equation}
Thus, the condition that accelerating expansion occurs is the same as
in the String Frame. It should be noted that the induced metric in
the String Frame and the Einstein Frame coincide with each other
for $p = 3$ because the dilaton (\ref{eq:dilaton}) is constant in this case.

\section{Discussion \label{section:discussion}}

In the previous section, we dealt with a simple situation that
a probe brane goes away from the neighborhood of the source branes
to infinity. If the probe brane approach the source branes, 
the scale factor decreases as the inverse of that in the previous section.
Then the other situations, for example, scattering and bound state of branes,
are easy to imagine. In the former case, the brane contracts first, 
then bounces and finally expands. In the latter case, the brane
continues to expand and contract periodically.

In this paper, we followed the dynamics of a probe brane, that is,
we neglected the back-reaction. This is justified if the probe brane
is light compared to the source branes. This means $N \gg 1$, which
we assumed in the analyses in section \ref{section:cosmology}. 
If $N \sim 1$, we have to treat both branes equally and the self-gravity of 
the branes must be taken into account \cite{DBSG1,DBSG2}.

Our analysis assumes a stability of the probe brane. There are possible 
instabilities due to brane bending and radiation from the brane 
\cite{Branonium}. Ref. \cite{Branonium} has given a preliminary analysis 
on such instabilities. They have found, for $p=6$, that the brane is stable 
classically against bending and that the radiation is dominated by the one 
into the bulk dilation field, which can be made sufficiently small 
by appropriate choice of the string coupling constant. 
One of the key assumptions they made is to treat the brane motion 
as non-relativistic one. In other words, their results have been obtained 
in the large-separation limit: $k/r^{7-p}\ll 1$ 
(or $r\gg r_g$ in our notation). We obtain some of the expressions 
for the scale factor for $r\ll r_g$ where the relativistic treatment is 
necessary in strict sence but we expect that the relativistic corrections 
do not change the stability discussed in \cite{Branonium}. 
Recently it was shown in \cite{stability} that time-variations in the
background moduli fields generally preclude the existence of stable
elliptical orbits.

Finally, although our study is based on the approximated Lagrangian
\ref{eq:Lagrangian}, it would be quite interesting and important to study
the exact Lagrangian. This will be our future work.

\section{Summary \label{section:summary}}

In this paper, we investigated the evolution of the scale factor
of a probe Dp-brane which move in the background of source Dp-branes.
When the probe brane move away from the source branes, it expands
by power law, whose index depends on the dimension of the brane. 
If the energy density of the gauge field on the brane 
is subdominant, the expansion is decelerating irrespective of 
the dimension of the brane. On the other hand, when the probe brane
is a Nambu-Goto brane, the energy density of the gauge field
can be dominant, in which case accelerating expansion occurs for $p \leq 4$. 
The accelerating expansion stops when the brane has expanded sufficiently 
so that the energy density of the gauge field become subdominant.
Although this is not the case with a probe D-brane, we could still obtain
tendency of high-energy effect of the Born-Infeld action.

The system which is investigated in this paper is too simple to be 
our universe. However, further investigation will give understanding for 
the relation between superstring theory and our universe.


\section*{Acknowledgments}

We would like to thank Eitoku Watanabe, Tetsuya Shiromizu, Kazuya Koyama,
Takashi Torii and Daisuke Ida for fruitful discussions. 
The work of KT is supported by JSPS.

\clearpage

\end{document}